# A Dynamic Model Of Cascades On Random Networks With A Threshold Rule

Daniel E Whitney[1]
*MIT, Cambridge MA 02139 USA*

Abstract

Cascades on random networks are typically analyzed by assuming they map onto percolation processes and then are solved using generating function formulations. This approach assumes that the network is infinite and weakly connected, yet furthermore approximates a dynamic cascading process as a static percolation event. In this paper we propose a dynamic Markov model formulation that assumes a finite network with arbitrary average nodal degree. We apply it to the case where cascades follow a threshold rule, that is, that a node will change state ("flip") only if a fraction, exceeding a given threshold, of its neighbors has changed state previously. The corresponding state transition matrix, recalculated after each step, records the probability that a node of degree $k$ has $i$ flipped neighbors after $j$ steps in the cascade's evolution. This theoretical model reproduces a number of behaviors observed in simulations but not yet reported in the literature. These include the ability to predict cascades in a domain previously predicted to forbid cascades without assuming that the network is locally tree-like, and, due to the dynamic nature of the model, a "near death" behavior in which cascades initially appear about to die but later explode. Cascades in the "no cascades" region require a sufficiently large seed of initially flipped nodes whose size scales with the size of the network or else the cascade will die out. Our theory also predicts the well known properties of cascades, for instance that a single node seed can start a global cascade in the appropriate regime regardless of the (finite) size of the network. The theory and simulations developed here are compared with a foundational paper by Watts which used generating function theory.

---

[1] Sr Research Scientist, Engineering Systems Division, MIT ESD, Room E40-243, 77 Mass Ave, Cambridge MA 02139

# I. INTRODUCTION AND MOTIVATION

## A. Motivation

Percolation phenomena on networks have been used to model cascades on networks for several situations: bond and site percolation on regular networks Petermann and de los Rios [1], Centola et al [2] and bond percolation on random networks Newman 2006 [3]. Calloway et al [4] assumed that the network was infinite and weakly connected so that generating functions could be used to derive a static condition under which the probability of a giant connected cluster approaches unity. This method was used by Newman, Strogatz, and Watts [5] to re-derive the Molloy-Reed [6] criterion for percolation. Watts [7] extended this method to the case where a node will not join the percolation cluster unless a certain threshold fraction of its neighbors has already joined, and uses this to model cascades with local dependencies. Watts' analysis is revisited here as a motivation for the present work.

The cascade problem is defined here as follows: One forms an approximate Erdös-Rényi undirected random network comprising $n$ nodes linked independently with probability $p$ so that the average nodal degree $z = np$. A threshold $\phi$ is defined and applied to all nodes, requiring that a node will change state (or "flip") if a fraction of its neighbors equal to or exceeding $\phi$ has previously flipped.[2] To start the process, a seed node or nodes is selected at random from the network and arbitrarily flipped from the *off* to the *on* state. Neighbors of flipped nodes are checked to see if their threshold has been exceeded and, if so, these are flipped on the next step, forming a cohort of newly flipped nodes. The process goes on in this way until no more nodes can be flipped. Either all or almost all of the nodes flip or else the process typically stops when only a small fraction of the network has flipped. There are no parameter settings, as there are in the SIR model (Newman (2002) [9], Volz [10]) in which the cascade can be tuned to stop at any chosen point short of completely flipping the network.

The generating function approach equates a cascade with the emergence of a giant cluster in an infinite network. It calculates the probability that a giant cluster emerges by calculating the average size of a tree-like cluster of nodes and finding the conditions under which this average becomes unbounded. It is thus a static analysis, an existence proof. Lopez-Pintado (2006) [11] uses a dynamic Markov model to predict cascades on infinite networks with a threshold and compares networks with different degree distributions. Tlusty and Eckmann [12] study a phenomenon called "quorum percolation," meaning that in addition to a threshold (in their case a number of flipped neighbors rather than a fraction) it is shown that an initial seed of arbitrarily flipped nodes must be large enough or else the cascade will die out. This minimum size is called the quorum. Jackson and Yariv [13] use mean field theory to derive conditions for a similar phenomenon they call "nettipping." Gleeson and Cahalane [14] and Gleeson [15] derive a cascade condition for Watts' problem by assuming the network is locally tree-like and without cycles. They use this condition to show that cascades will occur in the domain where Watts predicted that no cascades would occur if the seed is sufficiently large.

The goal of this paper is to provide a theory that expressly assumes that the network comprises a given finite number of nodes and may have any average nodal degree, not just a small one. The theory is dynamic, modeling a step-by-step process. It reproduces or generalizes many results from generating function theory as well as new ones such as the requirement for a quorum for certain combinations of $z$ and $\phi$.

The approach taken here departs in several ways from the typical network- or generating function-based approaches. First, the standard treatment assumes that the networks are tree-like and without cycles, limiting the average nodal degree to $z = 1$ or a little above. Here, we take any value such that $z > 1$. Second, derivations of percolation conditions must either calculate the degree distribution of each cohort of newly flipped nodes or assume one. Usually the distribution is assumed to be the same as that of the parent network. Here we acknowledge that it is different and must be calculated anew for each step in the cascade. Failure to do so results in large errors when the theory is compared to simulations. Third, network-based derivations must account for the difference between the "incoming" and "outgoing" edges of a newly flipped node, where "incoming" edges lead back to flipped nodes and "outgoing" edges link to unflipped ones. If, as in Calloway et al [4] or Newman (2006) [3], the network is assumed to be tree-like then there is always one incoming edge, and if there is no threshold then the node will flip. If, as assumed here, the network is not tree-like, then the number of incoming edges must be calculated for each node and depends on its degree. If, in addition, we model the threshold case, then the number of incoming edges required to flip a node depends on its degree and the threshold, and may be much greater than one. In the early stages of a cascade, the threshold forces newly flipped nodes to be of lower than average degree. The combination of extra required incoming edges and the threshold means that newly flipped nodes will have relatively few outgoing edges. The approach described below takes all of these factors into consideration.

---

[2] Lopez-Pintado (2008) [8] analyzed the case where nodes can unflip but in this paper we assume that nodes, once flipped, stay flipped.

Typical theory also derives the conditions for the occurrence of only an expanding cascade, in the sense that every cohort of flipped nodes is larger than the previous cohort. Contracting processes are assumed to die out. Here we show that the existence of an ever-expanding cohort is a sufficient condition because we derive conditions under which an initially contracting process will, perhaps after many steps, reverse and become permanently expanding. We call this the "near death" phenomenon and present a sufficient condition for it to occur.

### B. Outline of the Paper

The paper is organized as follows. Below we review Watts' analysis. In Section II we show simulation results for the full range of $z$ within both the cascade and no cascade regions of Watts' Figure 1, including three specific reproducible findings. In Section III we develop the theory that is intended to reproduce the three findings observed in the simulations, while in Section IV we compare theory and simulations with respect to these findings. The paper concludes with a discussion of the results.

### C. Watts' Cascade Model

In Watts' analysis, infinite Poisson random networks are characterized by their average nodal degree $z$ and threshold $\phi$. Initially all nodes are unflipped except for a randomly chosen "small" seed. We will employ Watts' alternate terminology $K^* = \lfloor 1/\phi \rfloor$, where $K^*$ is a threshold number of edges. Watts extended the theory in [5] to derive conditions in terms of $z$ and $K^*$ under which the average size of clusters of vulnerable nodes (nodes needing only one flipped neighbor in order to flip, equivalently, nodes having degree $k \in [1, K^*]$) diverges and thus that flipped vulnerable clusters comprise a finite fraction of the infinite network. Flipping a finite fraction of the infinite network defines a global cascade. This approach finds the conditions for global cascades that stay entirely inside clusters of vulnerable nodes, which are tree-like. Watts' theoretical result can be represented as a region in $K^* - z$ space, shown as dashed lines in **Figure 1**, inside of which global cascades are predicted to occur all or almost all the time, and outside of which they occur never or almost never. The black dots represent values of $z$ and $K^*$ for which an entire simulated finite network of 10000 nodes flips a small percent of the time starting from randomly selected single node seeds, defining an approximate boundary between the Global Cascades region and the No Global Cascades region.

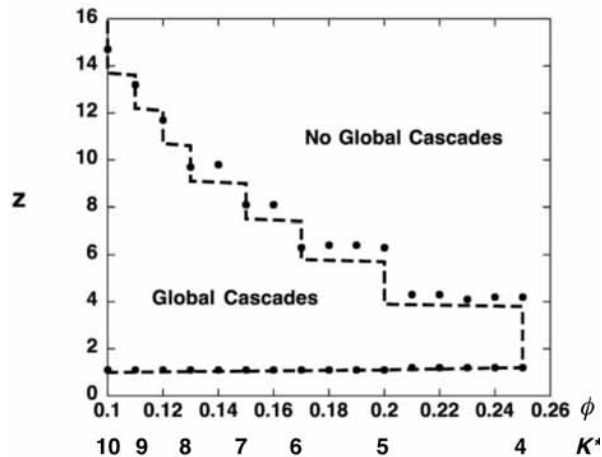

Figure 1. Watts' Cascade Region. (Adapted from Watts [7].) Theory based on assuming the network is infinite and weakly connected predicts that global cascades will occur all or almost all the time for values of $z$ and $K^*$ in the region inside the dashed lines after a "small" seed is flipped. Outside that region no global cascades should occur. A global cascade is defined as flipping one or more clusters of vulnerable nodes comprising a finite fraction of the infinite network. The black dots represent values of $z$ and $K^*$ where the whole network flips in perhaps 1% of attempts in simulations on networks with 10000 nodes.

Figure 1 shows that global cascades occur in the simulation well inside the region in which no global cascades are predicted to occur. The reason for this is summarized here and in more detail in Whitney [16], who observed many simulations closely. As reported by Watts [7], as one increases $z$ for a given $K^*$ starting just above $z = 1$, one creates networks with successively larger clusters of vulnerable nodes until in the middle of the Global Cascades region the bulk of the network comprises one large vulnerable cluster. A single-node seed easily can start a global cascade because most seeds are vulnerable or have several vulnerable neighbors. This mass of vulnerable nodes can flip any remaining stable nodes (those requiring more than one flipped neighbor in order to flip). As $z$ approaches the step-like upper boundary, however, the fraction of vulnerable nodes in the (finite simulated) network declines and they begin gathering in a larger number of smaller clusters. In the vicinity of the upper boundary, large cascades are almost impossible to launch, even though the network may contain as many as 20% vulnerable nodes, because nearly all of them are isolates. Whitney [16] A few occupy small clusters (typically 5 clusters comprising no more than 10 or 15 vulnerable nodes each in a network comprising 4500 nodes). A cascade that flips the whole network[3] can nevertheless emerge if a single-node seed happens to have a neighbor in one of these vulnerable clusters. Once any node in such a cluster flips, the rest of the cluster spontaneously flips. This event acts to multiply the size of the seed. Should two (or more) of these flipped vulnerable nodes have a common stable neighbor outside the cluster, it will flip if its threshold is exceeded. If, in addition, this newly flipped node has a neighbor in another vulnerable cluster, the cascade can effectively "hop" to that cluster and, under fortunate circumstances, more such hops occur and a global cascade emerges. In networks of the size Watts used in his simulation, cluster hopping can occur for values of $z$ that are in the No Global Cascades region in Figure 1. Networks with identical $z$ are more prone to cluster-hopping if they have larger vulnerable clusters. In Whitney [16] it is shown that cluster-hopping can occur with networks in a wide range of sizes from 2500 to 36000 nodes and there is no reason to expect that it would not occur for larger networks, although this was not investigated.

Cascades in the No Global Cascades region appear to violate the generating function theory. In addition, this theory does not predict any mechanisms such as cluster-hopping. These observations, plus the fact that real networks are finite (as are the simulations used to test theories) motivate the desire to have a theory that directly allows us to specify the size of the network and its average nodal degree and to be free of the restriction that the network be tree-like or that the analysis, as in Watts [7], be limited to tree-like subgraphs of a larger network that is not tree-like.

## II. Simulation Method and Results

In this section we present the simulation method and three findings from simulations of cascades on finite random networks with various values of $z$ and $K^*$ covering both the cascade and no cascade regions.

Random networks are realized using a simple Matlab® routine which takes $n$ and $p = z/n$ as inputs. The resulting degree distribution $p_k$ is checked for conformity to the required binomial form. Values of $K^*$ and seed size $|S|$ are chosen, and a simulation is launched by choosing the seed nodes at random. Hundreds of simulations are performed by choosing different identically-sized seed sets on the same network. Then a new network is realized and the process is repeated. No significant changes in behavior are noted between using the same network or changing to a different one aside from the effect of realizing a slightly different value of $z$. The outcome of each simulation is recorded in terms of number of nodes flipped on each step and total number flipped. If the whole or almost whole network flips, we say that a Total Network Cascade or TNC has occurred.

The first finding is shown in Figure 2.

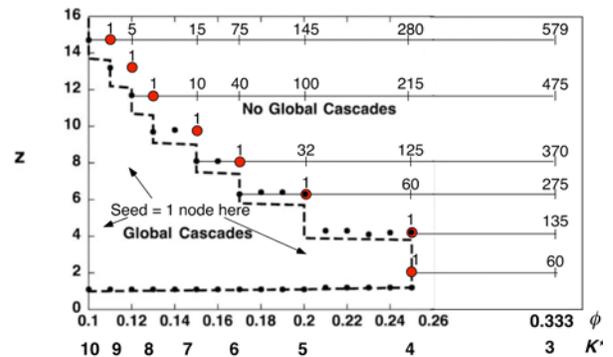

Figure 2. Simulation Results with Different Size Seeds. Here $n = 4500$. The large closed circles reproduce Watts' result, where seed size $|S| = 1$. In the region labeled "Global Cascades," a seed comprising one node can start a TNC. In the region labeled "No Global Cascades," larger $z$ and/or smaller $K^*$ require larger size seeds. Numbers on the line extending to the right of a large closed circle represent "transitional" (see the text for the definition) sizes of seeds needed to start a TNC. A clear pattern of contours of constant $|S|$ can easily be imagined from these data. (Figure adapted from Watts [7].)[4]

---

[3] Later in the paper we will call cascades that flip all or most of the nodes in a finite network Total Network Cascades (TNCs) to distinguish them from global cascades defined as flipping a finite fraction of an infinite network because the latter criterion does not apply to finite networks.

[4] It should be noted that only integer values of $K^*$ exist, giving rise to the individual values of $|S|$ shown in Figure 2. Additional values, such as the black dots at intermediate values of $\phi$, do not in fact represent different conditions because all these values of $\phi$ alias to the same value of $K^* = \lfloor 1/\phi \rfloor$. The stair-step representation of the boundary is thus something of an illusion but is carried on here because in the literature it

Numerical values in this figure agree with those derived by Gleeson and Cahalane[14].

This figure represents two cascade regimes in networks of 4500 nodes. In the Global Cascades region, and in the No Global Cascades region up to the black dots and large closed circles, a seed comprising a single node will start a TNC in a finite network of any size in the range tested (2500 – 36000 nodes). In the No Global Cascades region above and to the right of the dots, larger seeds, scaling with the size of the network, are needed. The values of seed size shown in this region are called "transitional" meaning that this seed size will start TNCs occasionally, perhaps 20% of the time. Somewhat smaller values will never start TNCs while values somewhat larger will start TNCs every time.

The second finding is that, in the No Global Cascades region, there is a fairly sudden transition in seed size from too small to large enough, typically spanning no more than about 10% of the seed's transitional value. A typical result is shown in **Figure 3**. This behavior is similar to a phase transition (i.e., there is a critical seed size below which there is almost surely no global cascade, and above which there is almost certainly a global cascade).

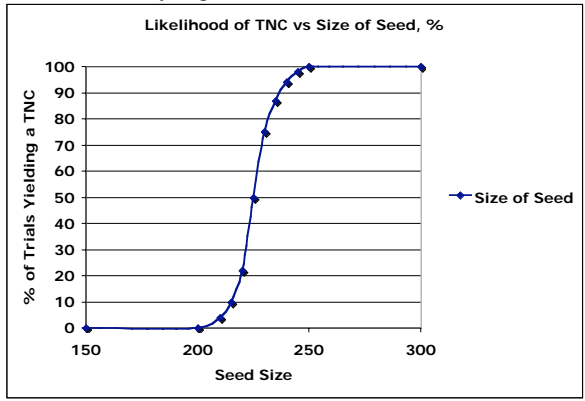

Figure 3. Likelihood (in %) of a TNC vs Size of Seed When $n = 4500, z = 11.5,$ and $K^* = 4$. The transition value of seed size is ~215, a value at which TNCs occur perhaps 20% of the time.

The third finding is that, in the region where $|S| > 1$, cascades can become TNCs even though they begin by contracting. Figure 4 shows several trajectories, tracking the number of nodes flipped on each step, launched by different values of $|S|$ corresponding to different points in the No Global Cascades region in Figure 2. The transition from "no

TNCs" to "100% TNCs" is clearly seen. For values of $|S|$ below the transition value, the number flipped on each step falls more or less monotonically and finally goes to zero. For values of $|S|$ in the transition region, using the same network and a different randomly chosen seed of the same size, a TNC may or may not occur. This is illustrated by the two trajectories corresponding to $|S| = 220$. Here, because the size of the seed is in the transition region, the number of nodes flipped per step falls at first and languishes near zero, and usually terminates without a TNC. But occasionally it begins to grow rapidly later, and a TNC results. We call this the "near death" phenomenon and associate it with some kind of critical mass (other than the minimum seed size) which will be explained in Section IV.

Trajectories in the transition region vary greatly in duration, regardless of whether they eventually become TNCs or not. For increasing values of $|S|$ above the transition value, the trajectory initially falls less and for fewer steps, and then rises faster. If a large enough seed is chosen, the trajectory rises immediately. These trajectories have been terminated as soon as more than 10% of the network has flipped but if left to continue will terminate when about 90% of the network has flipped. The running total of flipped nodes follows a sigmoid or logistic curve typically predicted or observed for a range of phenomena from population growth to the diffusion of innovations. (Bass [19] Griliches [18] Rogers [17] Valente [20]) An example appears in **Figure 7**.

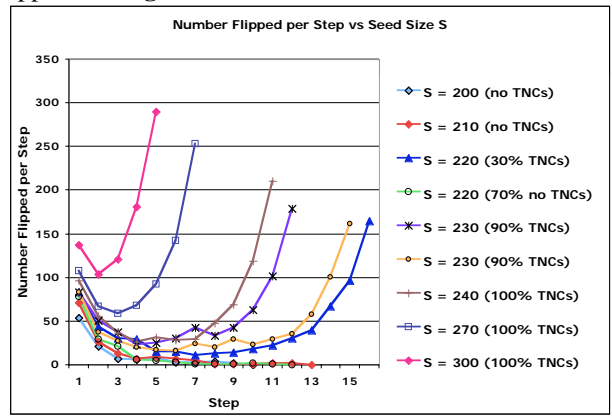

Figure 4. Typical Behavior of a Cascade for Different Values of $|S|$, the Size of the Seed. Here, $n = 4500, z = 11.5,$ and $K^* = 4$. Each trajectory plots the number of nodes flipped on each step. Trajectories for $|S| = 270$ and $|S| = 300$ are averages of 10 simulations each. Other trajectories are

---

has become customary to represent this boundary as a stair-step.

individual examples. Individual examples where $|S|$ is in the transition region vary a great deal from one another for the same conditions, as illustrated by the two examples corresponding to $|S| = 220$ and $|S| = 230$, respectively.

### III. Theoretical Model

The goal of the theory is to predict the average number of nodes flipped on each step as well as their degree distribution. From this information we seek to reproduce all the findings from simulations discussed previously, plus those found by Watts and Gleeson.[5] The theory assumes that the network is random and its degree distribution is binomial. The order in which the neighbors of a node flip does not matter, but rather only the current percent of flipped neighbors, in determining the probability that a node will flip. Each node is assumed to have the same threshold $\phi$, represented by $K^* = \lfloor 1/\phi \rfloor$. The theory therefore does not have to remember, when evaluating the next step, which nodes have how many flipped neighbors but rather only the probability that a node of given degree has a given number of flipped neighbors, based on how many nodes on average were predicted to flip on the previous steps.[6] The simulation, however, remembers all these items explicitly.

The derivation begins with the method for determining the distribution $p_k$ of the degree of an unflipped node, in a binomial random network of size $n$, $i$ of whose $k$ edges link to nodes that are in a subset comprising $S$ flipped seed nodes selected at random from the network while the remaining $k-i$ edges link to nodes that are among the remaining $n-S-1$ unflipped nodes.[7] This event may be expressed as the sum over $i$ of the product of two independent events $p_1(i,S)$ and $p_{nS}(k-i, n-S)$ where

**Equation 1**

$$p_1(i,S) = \binom{S}{i} p^i (1-p)^{S-i}$$

and

**Equation 2**

$$p_{nS}(k-i, n-S) = \binom{n-S-1}{k-i} p^{k-i} (1-p)^{n-S-1-(k-i)}$$

Then, the resulting degree distribution for an unflipped node is[8]

**Equation 3**

$$p_k = \sum_{i=0}^{k} \binom{S}{i} p^i (1-p)^{S-i} \binom{n-S-1}{k-i} p^{k-i} (1-p)^{n-S-1-(k-i)}$$

where $i \le \min(k,S)$.
Vandermonde's identity[9]

$$\binom{n}{k} = \sum_{i=0}^{k} \binom{S}{i} \binom{n-S}{k-i}$$

may be used to verify that Equation 3 is equivalent to the degree distribution of nodes in a binomial network

$$p_k = \binom{n-1}{k} p^k (1-p)^{n-1-k}$$

Multiplying $p_k$ in Equation 3 by $n-S$, the number of unflipped nodes, will give us the average number of unflipped but potentially flippable nodes of each degree $k$.

The decomposition represented by Equation 3 can be extended recursively, and this too will be useful in subsequent steps of the derivation.

**Figure 5** shows diagrammatically the first step captured in Equation 3 and the second step captured by Equation 5.

---

[5] Gleeson [15] handles the case where the seed comprises a single node separately from the case where the seed is a given fraction of the network's size. Our theory handles both cases with the same formulation.

[6] That is, even though the theory predicts the average degree distribution of flipped nodes, it uses only the number of flipped nodes to predict what will happen on the next step. This simplification is permitted by the assumption that the network's degree distribution is binomial. The limitations of this assumption are discussed later in the text.

[7] When there is no possibility of confusion, we will drop the terminology $|S|$ for the size of a set of nodes and simply refer to subsets of the network and the number of nodes in them using the same symbol, such as $S$ or $n-S$.

[8] This equation gives the same numerical results as equation (4) in Cohen et al [21] as well as the corresponding equation (not numbered) in Lopez-Pintado (2006) [11]. But their equations cannot be used in the given form for subsequent steps in a finite network.

[9] http://en.wikipedia.org/wiki/Choose_function

The value of $K^*$ divides the network into $k-$classes *vulnerable*, *first order stable*, *second order stable*, etc., as shown in **Table I**. Vulnerable nodes flip if they have one or more flipped neighbors, first order stable nodes flip if they have two or more flipped neighbors, etc.

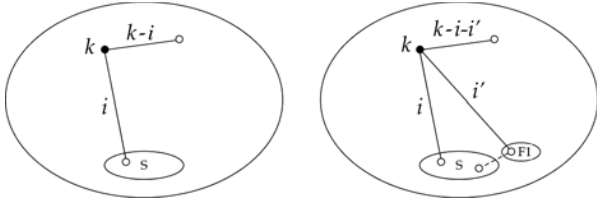

Figure 5. Left: Illustrating the first step in the cascade where a randomly selected unflipped node with degree $k$ in the network (the large oval) has $i$ links to flipped nodes in the seed set $S$ (the small oval) and its remaining $k-i$ links to unflipped nodes in the rest of the network. A closed circle is a single node while an open circle stands symbolically for one or more nodes. Some of the unflipped nodes will flip, forming set $F1$. Right: The same case for the second step of the cascade, where the seed set has flipped additional nodes $F1$. Here a randomly selected unflipped node with $k$ links has $i$ links to $S$, $i'$ links to $F1$, and its remaining $k-i-i'$ links to unflipped nodes elsewhere in the network. Some of these nodes will flip, forming set $F2$ (not shown). Sets $S$ and $F1$ are shown as distinct because both theory and simulations are designed to not count possible overlaps; these overlaps can arise because nodes in each of these sets have neighbors in the respective sets that have already flipped and thus should not be counted multiple times. The dashed line between nodes in $F1$ and $S$ represents the links from $S$ that flipped nodes in $F1$. These links cannot be included in the $i'$ links that are available to flip nodes on the next step. This fact is reflected in the formulation.

Table I. Formulae for the Average Number of Nodes of Different Stability Categories or *k*-classes in the Whole Network and in the Seed. Only the first three categories are shown because the pattern repeats in an obvious way.

| *k*-class of Node | Degree range | Average number in the network | Average number in $S$ |
|---|---|---|---|
| Vulnerable | $k \in [0, K^*]$ | $n \sum_{k=0}^{K^*} p_k$ | $S \sum_{k=0}^{K^*} p_k$ |
| First order stable | $k \in [K^*+1, 2K^*]$ | $n \sum_{k=K^*+1}^{2K^*} p_k$ | $S \sum_{k=K^*+1}^{2K^*} p_k$ |
| Second order stable | $k \in [2K^*+1, 3K^*]$ | $n \sum_{k=2K^*+1}^{3K^*} p_k$ | $S \sum_{k=2K^*+1}^{3K^*} p_k$ |

Using the information in Equation 3 and **Table I** we can determine the probability that an unflipped node of degree $k$ has $i$ links to $S$ (called being "hit" $i$ times) on step 1 and consequently whether it should flip or not. For example, vulnerable nodes will flip if $i \geq 1$, first order stable nodes will flip if $i \geq 2$, etc.

The result of the first step is that a new set of flipped nodes is created, which we call $F1$, comprising $|F1|$ nodes. The next step in the cascade flips additional nodes which have links to $F1$ and/or to $S$. Flipping can occur two ways: a node may flip because it has not been hit before but has enough links to $F1$ to make this happen, or it may already have some links to $S$ and the additional links to $F1$ are sufficient to make it flip. Concurrently, some nodes will gain links to $F1$ that are insufficient to make them flip, whether or not they have links to $S$. Finally, some nodes will remain unhit. The newly flipped nodes are called $F2$ and there are, by our convention, $F2$ of them.

Formally, extending the notation of **Equation 1** and **Equation 2** we can write the degree distribution of unflipped nodes in the network outside of $S$ and $F1$ at this point as

**Equation 4**

$$p_k = p_1(i, S) p_2(i', F1) p_{nSF}(k - i - i', n - S - F1)$$

or, equivalently

**Equation 5**

$$p_k = \sum_{i=0}^{k} \sum_{i'=0}^{k-i} \binom{S}{i} p^i (1-p)^{S-i} \binom{F1}{i'} p^{i'} (1-p)^{F1-i'}$$
$$\binom{n-S-F1-1}{k-i-i'} p^{k-i-i'} (1-p)^{n-S-F1-1-(k-i-i')}$$

where $i \leq S$, $i' \leq F1$, and $i + i' \leq k$.

Formally, we can proceed to find $F2$ as we did to find $F1$ in the first step, determining the average number of vulnerable, first order stable, second order stable, etc. nodes that have flipped or been hit a given number of times up to and including the second step by multiplying these expressions by $n - S - F1$, the number of nodes in the network outside of the ones that flipped before. But this will not give us accurate answers, for the following reasons.

The sub-network $F1$ is not in fact representative of the rest of the network, since it consists of newly flipped nodes. These nodes necessarily have fewer edges than typical nodes, owing to the relative ease with which they are flipped compared to nodes of higher degree. In fact, we observe in theory and in typical simulations that the average nodal degree $z_{F1o}$ of $F1$ is as little as 70% of that of the original network. The rest of the network $n - S - F1$ then has somewhat higher average nodal degree than the original network by a few percent. Furthermore, as noted above, nodes in $F1$ have one or more links to nodes in $S$ which are thus unavailable to flip new nodes. This further reduces the effective average nodal degree of nodes in $F1$. Thus Equation 5 must be rewritten as Equation 6 to reflect this, and the distinctions mentioned are recognized by attaching appropriate subscripts to the probabilities, which must be calculated to suit, using the calculated degree sequences of $F1$ and $n - S - F1$.

**Equation 6**

$$p_k = \sum_{i=0}^{k} \sum_{i'=0}^{k-i} \binom{S}{i} p^i (1-p)^{S-i} \binom{F1}{i'} p_{F1}^{i'} (1-p_{F1})^{F1-i'}$$
$$\binom{n-S-F1-1}{k-i-i'} p_{nSF1}^{k-i-i'} (1-p_{nSF1})^{n-S-F1-1-(k-i-i')}$$

Here, the relevant parameters, reflecting the reduced average nodal degree and unavailable edges of flipped nodes, are

**Equation 7**

$p_{F1} = z_{F1}/n$

$z_{F1} = \dfrac{F1 z_{F1o} - \# links\_to\_S}{F1}$

where $z_{F1o}$ = average nodal degree of $F1$

$\# links\_to\_S = \sum_{\lambda=1}^{\infty} \lambda F1_{\lambda}$

where $\lambda$ = a number of links from $F1$ to $S$

$F1_{\lambda}$ = the number of nodes in $F1$ that have $\lambda$ links to $S$

**Equation 8**[10]  $p_{nSF1} = p \dfrac{n - S - \alpha_{F1} F1}{n - S - F1}$

$\alpha_{F1} = z_{F1o}/z$

---

[10] This equation is derived in the Appendix.

To generate subsequent average numbers of hits on nodes in the various $k-$classes on later steps, it is convenient to define

**Equation 9**  $_i h_j = p_1 p_2 ... p_j$

where

$_i h_j$ = *the probability of being hit i times after j steps*,

$i = 0 : m$ hits, $j = 1 : s$ steps

and $p_1$, $p_2$, etc. are defined in Equation 4, Equation 5, and their recursive successors.

Then we can write

**Equation 10**

$_i h_j = {_i h_{j-1}} \overline{F}_j + {_{i-1} h_{j-1}} 1F_j + {_{i-2} h_{j-1}} 2F_j + ...$

where $\overline{F}_j$ means the probability that an unflipped node is not hit by any nodes in $F_j$ and $qF_j$ means the probability that an unflipped node is hit $q$ times by nodes in $F_j$. Then the following Markov recurrence model for $_i h_j$ can be written as

**Equation 11**

$$\begin{bmatrix} _0 h_j \\ _1 h_j \\ _2 h_j \\ ... \\ _m h_j \end{bmatrix} = \begin{bmatrix} _0 h_{j-1} & 0 & 0 & ... & 0 \\ _1 h_{j-1} & _0 h_{j-1} & 0 & ... & 0 \\ _2 h_{j-1} & _1 h_{j-1} & _0 h_{j-1} & ... & 0 \\ ... & ... & ... & ... & ... \\ _m h_{j-1} & _{m-1} h_{j-1} & _{m-2} h_{j-1} & ... & _0 h_{j-1} \end{bmatrix} \begin{bmatrix} \overline{F}_j \\ 1F_j \\ 2F_j \\ ... \\ mF_j \end{bmatrix}$$

where typical initial conditions are

**Equation 12**

$_0 h_0 = \overline{S} = (1-p)^S$

$_1 h_0 = 1S = Sp(1-p)^{S-1}$

$\overline{F}_j = (1 - p_{F_{j-1}})^{F_{j-1}}$

$1F_j = F_{j-1} p_{F_{j-1}} (1 - p_{F_{j-1}})^{F_{j-1}-1}$

$2F_j = F_{j-1}(F_{j-1}-1) p_{F_{j-1}}^2 (1 - p_{F_{j-1}})^{F_{j-1}-2}$

$etc.$

To use this model to find the average number of flipped nodes, we note that vulnerable nodes will flip if hit any number of times on step $j$, given that they were never hit before. Using entries from Equation 11, we can write

**Equation 13  Vulnerable nodes flip if:**

Hit once ($i = 1$): $_1h_j = {_0h_{j-1}}F_j$

Hit twice ($i = 2$): $_2h_j = {_0h_{j-1}}2F_j$

Hit thrice ($i = 3$): $_3h_j = {_0h_{j-1}}3F_j$

*etc.*

This is equivalent to rewriting Equation 11 using only the diagonal of the matrix, and setting all other entries to zero including the first row.  Note that this formulation assumes that if a node was hit $i$ times on step $j$ then it was hit no more than $i-1$ times on any previous step, and that the number of hits never decreases.

First order nodes flip if they are hit two or more times on step $j$ and have not been hit more than once on any previous step.  Then we may write

**Equation 14  First order stable nodes flip if:**

Hit twice ($i = 2$): $_2h_j = {_1h_{j-1}}F_j + {_0h_{j-1}}2F_j$

Hit thrice ($i = 3$): $_3h_j = {_1h_{j-1}}2F_j + {_0h_{j-1}}3F_j$

Hit four times ($i = 4$): $_4h_j = {_1h_{j-1}}3F_j + {_0h_{j-1}}4F_j$

*etc.*

This is equivalent to writing Equation 11 using only the diagonal and the first sub-diagonal of the matrix, setting all other entries to zero including the first two rows. For second order stable nodes, we use the diagonal and the first two sub-diagonals, setting all other entries to zero including the first three rows. This pattern repeats for higher order stable nodes. Using Equations 10 - 14 in Equation 6, Equation 7, and Equation 8 and their recursive successors for successive values of $j$, we can generate time series for the average number of nodes in any category and $k-$class, such as first order stable nodes flipped, second order stable nodes hit twice, fourth order stable nodes not hit, etc., and we can keep track of the average hit and flip history of every $k-$class or nodal degree of node for comparison to simulations.

The theory is implemented in Matlab, as are the simulations.

## IV. Comparison of Theory and Simulations

### A. Prediction of TNCs in the Global Cascades Region

Inside the Global Cascades region, the theory predicts the average number of nodes flipped per step and predicts TNCs if the seed is one node, regardless of the size of the network within the range simulated (2500 – 36000 nodes).[11]  **Figure 6** shows example results, which agree well with the simulations reported in **Figure 2**.  The whole network does not flip, the reason for which is discussed below. Below the lower threshold boundary of **Figure** 2, the theory predicts that a fractional node will flip and the number of flipped nodes will remain small.  The theory knows only the probability that two nodes are connected and has no concept of vulnerable clusters or connectedness of the network as a whole.  Thus the theory effectively predicts that no TNC will occur for $z < 1$.  When $z$ reaches the upper boundary given by simulations for a given $K*$, the theory again predicts that a fraction of a node will flip and the total will stay small.  In order for the theory to predict a TNC here and above, the size of the seed must be increased.  The required size scales with the size of the network.  This behavior is discussed next.

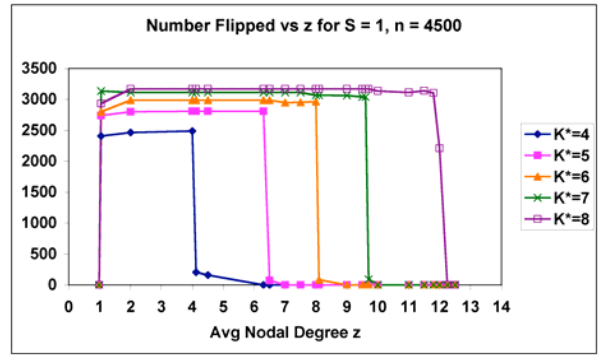

Figure 6. Behavior of the Theory in the Global Cascades Region.  The figure comprises vertical cross sections through Figure 2 for various values of $K*$. The seed is a single node in each case. For $z < 1$, the theory predicts that nothing will happen, but for $z$ just slightly above 1 a TNC occurs.  TNCs occur for larger values of $z$ until the upper boundary determined in simulations is reached.  Above these boundary values, larger seeds are needed to achieve TNCs.  The agreement between theory and simulations is very good.

### B. Prediction of TNCs in the No Global Cascades Region

In cases where the size of the seed is below (see **Figure 13** left) or above (see **Figure 7**) the transition region, agreement is quite good.  When the seed is in the transition region, the theory's cascades are usually less energetic than the simulations' and usually take longer to flip large numbers of nodes on each step.

---

[11] As the network becomes larger, it takes more steps for the TNC to emerge.

**Figure 7** reproduces the sigmoid shape commonly observed and modeled by various researchers in the field of diffusion of innovations and other growth phenomena. Agreement between theory and simulations is good up to about step 12. The theory does not flip the whole network. Two reasons have been identified.

The first is that the theory assumes that the flipped and unflipped subnetworks are binomially distributed, which is not true. See Figure 8 for an example. Instead, both gradually deviate from binomial as the cascade proceeds. When the unflipped part gets quite small the deviation from binomial becomes extreme, even in the simulations. It may be that the theory, based on assuming that both subnetworks are binomial with evolving values of $p$, simply cannot reproduce the simulation near the end of the TNC.

The more likely reason is that the theory, as noted above, takes no account of the fact that the network is connected and is based only on the probability that any randomly selected pair of nodes is connected. The theory sees the network merely as an array comprising the degree distribution of the unflipped nodes at any step. This is a fundamental characteristic of this theory. **Figure 9** shows that nearly all the vulnerable and first order stable nodes flip in a typical TNC, indicating that enough of these nodes are likely to get hit the requisite number of times, but that nodes requiring 2 or more hits are less likely to get enough. **Figure 10** shows that the most likely event on a given step is that no nodes are hit at all even when the number flipped on the previous step is large, and that most events comprise one or two hits on a given node. **Figure 11** shows the result: Nodes needing to be hit two or more times are much less likely to achieve this than nodes needing fewer. In a simulation, connectivity would take care of flipping the remaining nodes.

In addition, the theory has no way to predict how many steps the cascade will last, so it has no way to ensure that enough steps will occur in order that the probability of a node being hit enough times will sum to unity. It is evident from **Figure 9** that this probability in general sums to less than unity, but no explicit calculation of this probability has been made.

Finally, if this surmise is correct, we should observe more nodes flipping in a TNC if $K*$ is larger because this increases the number of vulnerable and first order stable nodes relative to the rest. Indeed we do observe this, as illustrated in **Figure 6**. This is also observed in the No Global Cascade region but is not shown.

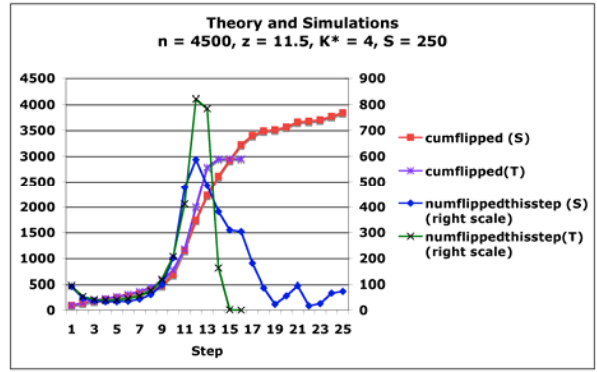

Figure 7. Complete TNC Comparing Theory (T) and One Example Simulation (S). Agreement is good up to about step 12 or 13, at which point the theory appears to lose the ability to flip more nodes. Even the simulation fails to flip every node. (The total flipped does not include the 250 nodes in the seed, so in fact the simulation flips 250 more than shown in the figure.)

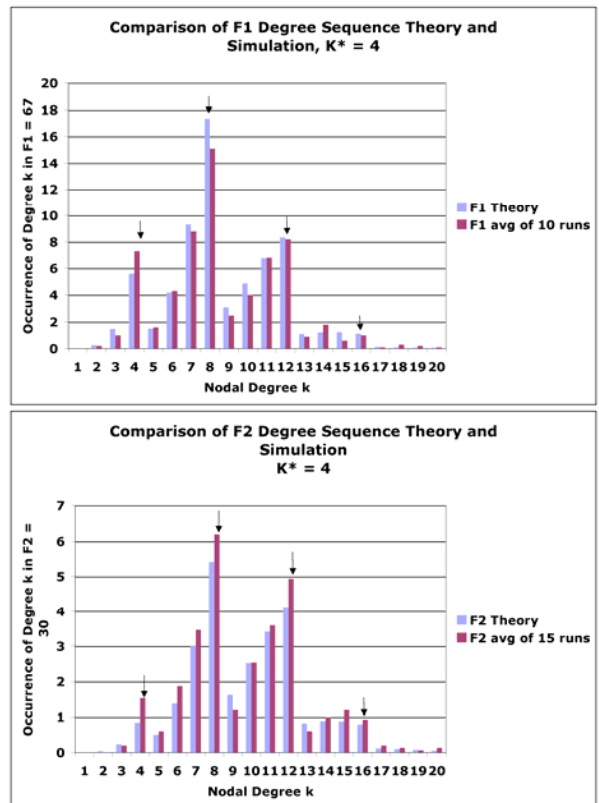

Figure 8. Comparison of Predicted and Simulated Degree Sequences of Flipped Nodes, First and Second Steps. Here $n = 4500, z = 11.5, S = 215, K* = 4$. Both theory and simulation display a jagged profile with peaks at multiples of $K* = 4$ (marked by arrows). Theory and simulations agree well but not perfectly.

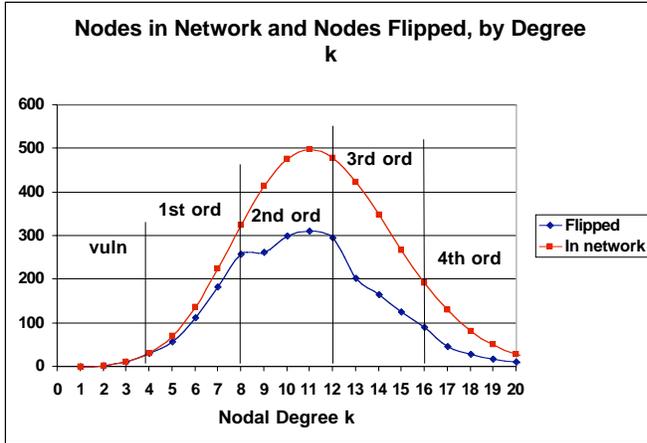

Figure 9. Example Comparison of Number of Nodes Flipped in a TNC with the Total Number in the Network, by Nodal Degree. Here, $n = 4500, z = 11.5, K^* = 4$. The graph is divided into the k-classes vulnerable, first order stable, etc., where vulnerable nodes need one flipped neighbor to flip, first order nodes need two, etc.

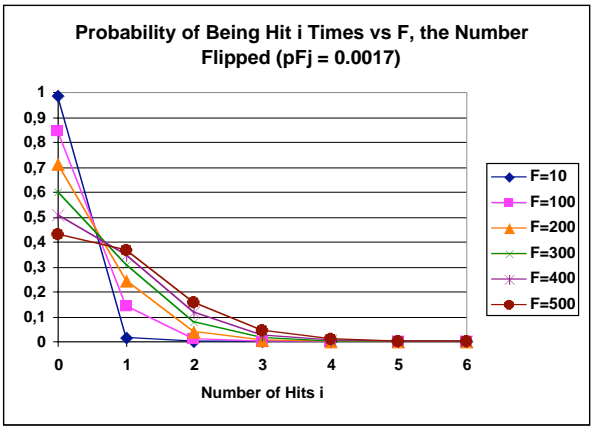

Figure 10. Probability of Being Hit $i$ Times by a Flipped Set Comprising $|F|$ Nodes. Near the end of a TNC, $|F|$ can reach into the hundreds $(n = 4500, z = 11.5, K^* = 4)$. Even then, the likelihood of being hit more than twice on a given step is low compared to being hit once or not at all.

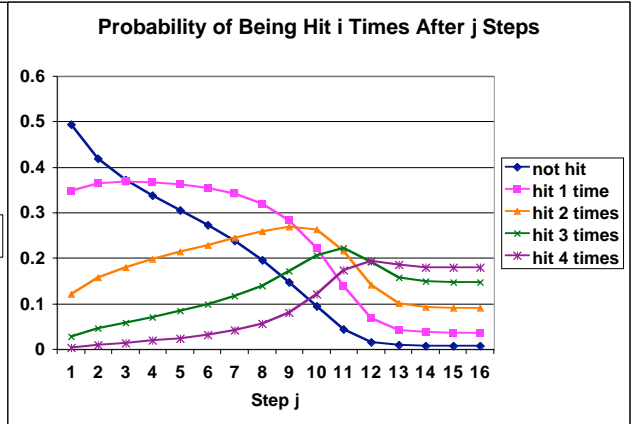

Figure 11. Example Evolution of Probability of Being Hit $i$ Times after $j$ Steps. $n = 4500, z = 11.5, K^* = 4$. The likelihood that more than 2 hits have accumulated on a node is low compared to fewer hits. Thus nodes of second order or higher are less likely to flip than vulnerable or first order stable nodes.

The sequence of flipping does not proceed by flipping the vulnerable nodes first, then the first order stable, then the second, and so on, but instead a TNC involves nodes from several of the lower k-classes right away (see Figure 8) and later brings in the higher ones. Even at the end, nodes from all k-classes are flipping.

### C. Prediction of Transition Seed Size in the No Global Cascades Region

To assess the accuracy of the theory in the No Global Cascades region, we use four examples for which extensive simulation data have been generated. These cases are listed in Table II. Each case is denoted by its value of $K^*$ and transitional value of $|S|$. Simulation data were generated by creating a random network with $n$ nodes with the specified $z$, setting the corresponding value for $K^*$, picking at random a seed of size $|S|$ at, above, or below the transitional value, letting the process evolve step by step, and recording a variety of data (number of nodes flipped on each step, number of hits accumulated on unflipped nodes on each step, values of $z_{Fj}$, and so on), repeating this 100 times. Mean and standard deviation over these 100 runs were recorded. A new network was then created and this process repeated, altogether 10 times, resulting in about 1000 individual runs for each case.

Table II. Cases Studied for Comparison to Theory. Networks with other sizes in multiples of 4500 were also tried, with results that scaled directly with size but displaying no qualitative differences.

| Case | $n$ | $z$ | Transitional $|S|$ | $K^*$ |
|---|---|---|---|---|
| 1 | 4500 | 11.5 | 215 | 4 |
| 2 | 4500 | 8 | 121 | 4 |
| 3 | 4500 | 11.5 | 90 | 5 |
| 4 | 4500 | 8 | 32 | 5 |

Using Equation 11 and Equation 12 and others like it systematically for different values of $S, z,$ and $K^*$ we can determine if the theory predicts a TNC or not. This prediction is binary in the sense that either the successive calculated values of $F_j$ will tend to zero or the total flipped will grow, approaching the size of the network. **Figure 12** shows, for the four cases defined in Table **II**, that the theory predicts that the transition from no TNCs to TNCs occurs within or at the upper end of the range of seed sizes over which the actual transition occurs in the simulations, showing that the theory can predict the transitional seed size within or close to the correct band of values.

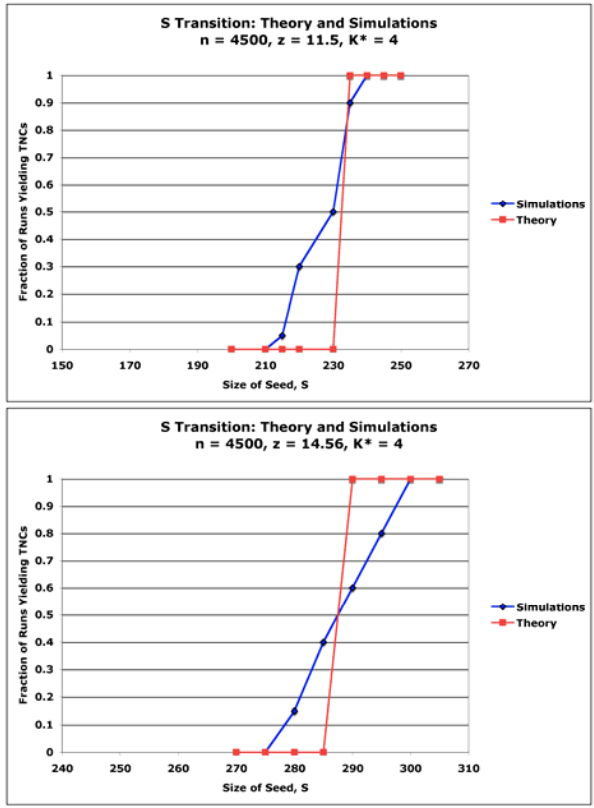

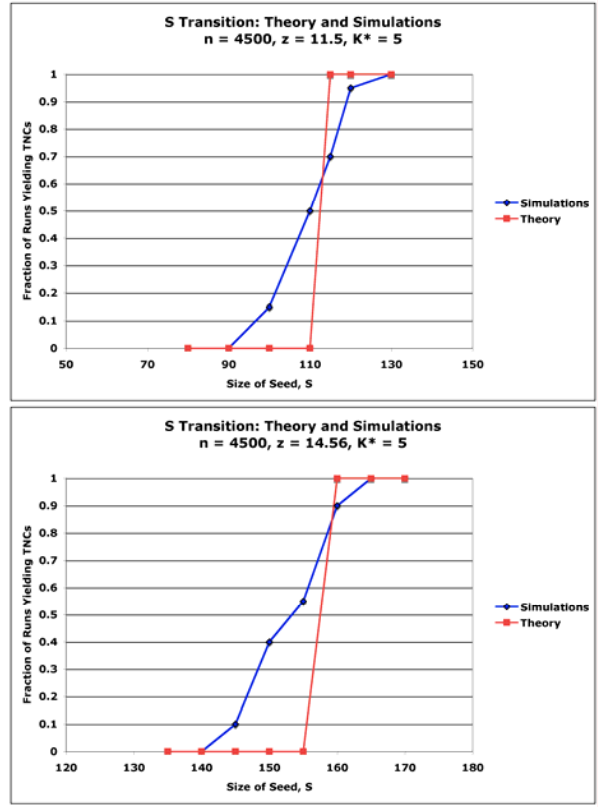

Figure 12. Comparing the Ability of the Theory to Predict TNCs vs $|S|$ to Simulations for Four Cases. The theoretical prediction is binary while the simulation's prediction is probabilistic. The theory predicts that TNCs will occur within or at the upper end of the range of values of seed size over which the probability of a TNC rises from zero to one in simulations.

### D. "Near Death" Phenomenon

The theory predicts that cascades launched with a transition value of $|S|$ will, on average, terminate without causing a TNC, but occasionally a TNC occurs in the simulation. The contrast is shown in **Figure 13** where $|S| = 215$, a transitional value, applies to both cases. The reason for this is almost certainly due to variation in the results of each step in the simulation. Such variations are often huge (30% to 50% differences in the number flipped on a given step from one simulation to the next using a different seed of the same size) and can occasionally allow a cascade launched from a seed of, say, 215 nodes to be able to flip 20 or more nodes on the 4[th] step, a level of flipping that is average for a successful TNC launched with $S = 235$ nodes. Thus seeds with sizes in the transition region of 215 nodes can succeed occasionally due to lucky variations, and it is this dependence on variations that gives rise to the probability of a TNC in the transition region being between zero and one. It appears that something changes or some threshold is exceeded in the successful ones. In the spirit of the literature on diffusion of innovations (Rogers [17]), we call this a critical mass phenomenon. In this section we discuss what the cause might be.

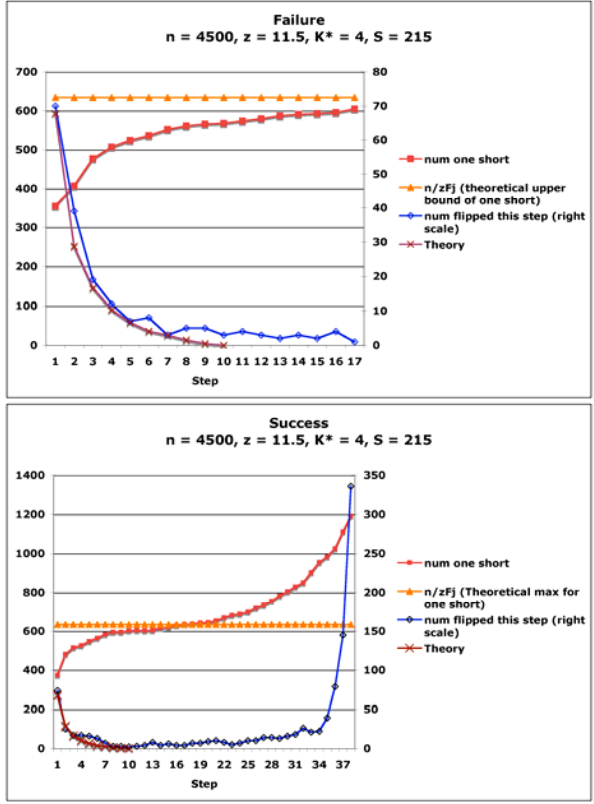

Figure 13. Comparing Identical Cases, One Where a TNC Occurs, the Other No TNC. Left: No cascade occurs, which is typical for this set of conditions. Theory and simulation agree well. Right: A rare cascade for the given conditions occurs, while the theory again predicts failure. The variables $num\_one\_short$ and $n/z_{Fj}$ in each panel are discussed in the text.

In the theory of diffusion of innovations, the "critical mass" is thought to be the number of initial innovators. Certainly this can be related to the minimum seed size needed to start a TNC. But in addition we observe in our simulations a second critical mass that causes a marginally sized seed to eventually succeed. Our hypothesis is that the number of nodes lacking one flipped neighbor is the pivotal parameter. We begin by observing that when the cascade is languishing and could stop, the number of nodes flipped on each iteration is small, perhaps as few as one or two nodes. In such a situation, the likelihood that a node in the network could have two or more neighbors in the set of newly flipped nodes is essentially zero. (This is illustrated in **Figure 10**.) Thus each unflipped node will be hit at most once on this step. If that node needs only one more hit to flip, it will flip. Nodes needing two or more hits to flip will have no chance of flipping. Thus we may safely confine ourselves to nodes needing only one more hit in order to flip. Let the number of these nodes be called $N_{os}$, where the subscript stands for "one short."

Let $q$ be the probability that a node in the network has a link to the newly flipped nodes on step $j$, called $Fj$ and comprising $F_j$ nodes:

**Equation 15** $$q = \frac{Fj\, z_{Fj}}{n}$$

where $z_{Fj}$ is the effective average nodal degree of $Fj$, which is defined in Equation 7. $Fj\, z_{Fj}$ is the average number of edges leading from the nodes $Fj$ and is thus the average number of nodes in the rest of the network that will have links to $Fj$. $N_{os}$ of these are one short. Thus the average number of nodes one short that have links to $Fj$ is

**Equation 16**
*average number of nodes one short with links to* $F_j = \dfrac{N_{os}\, F_j\, z_{Fj}}{n}$

This is the average number of nodes that $Fj$ is expected to flip. In order for the cascade to be at least self-sustaining, this number should equal $Fj$. If we let $Fj = 1$ then we want one additional node to flip for each member of $Fj$. Then Equation 16 becomes

**Equation 17** $$1 = \frac{N_{os} z_{Fj}}{n}$$

so that the minimum size of $N_{os}$, our critical mass, is

**Equation 18** $$N_{os} = n / z_{Fj}$$

For $n = 4500$, $z = 11.53$, $S = 215$, and $K^* = 4$, we can calculate $z_{Fj} \approx 7.25$, and we find that $N_{os} \approx 620$. For $z = 14.369$, $S = 280$, $K^* = 4$, we calculate $z_{Fj} \approx 9.5$, and we find that $N_{os} \approx 475$. The predicted number one short may be calculated from Equation 11 using logic similar to that in Equation 13 or Equation 14.

In Figure 14 we show the number of nodes one short at the moment that failed cascades died out (called max one short or $max\ N_{OS}$) for a range of seed sizes, comparing theory and simulations. The results are averages of 20 runs for each seed size. The agreement is good, and both the predicted and actual $max\ N_{OS}$ do not exceed the theoretical minimum required for a TNC until the simulations show a high likelihood that TNCs will occur. Results for the other cases in Table II are similar and are not shown. These findings indicate that the predicted upper bound is a good one in the sense that almost all processes that fail to exceed it die out while almost all processes that exceed it go on to become TNCs.

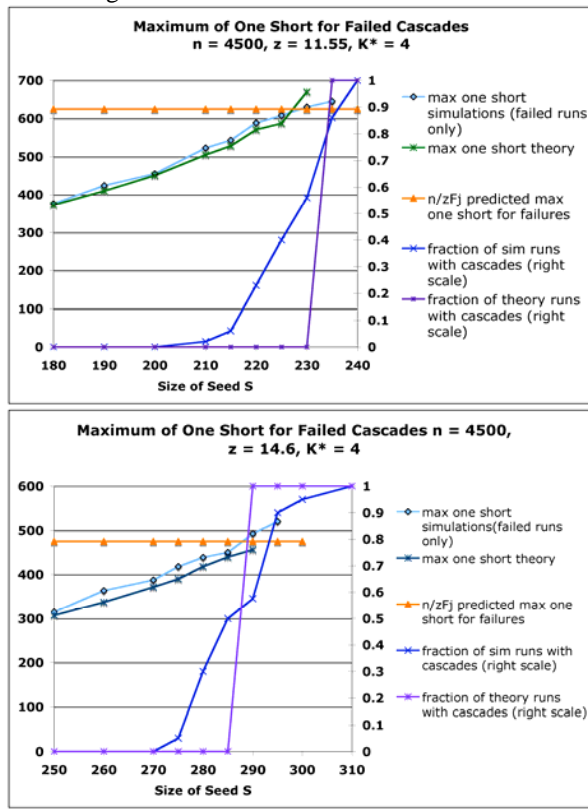

Figure 14. Comparison of Predicted and Actual Maximum Number of $N_{OS}$ for Failed Cascades. The results present averages of 20 runs for each seed size.

## V. Discussion and Conclusions

In this paper we formulated a dynamic theory of cascades on random networks with a threshold that treats the network as finite, and compared this theory with simulations. This theory is reasonably accurate and reproduces three identified phenomena observed in simulations. The theory does not in fact predict directly that a TNC will occur but instead predicts how many nodes on average will flip on each step. These probabilities evolve according to a recursive Markov model whose transition matrix must be recalculated at each step. For a given average nodal degree $z > 1$, we can predict the necessary size of seed in both the regions labeled by Watts as "Global Cascades" and "No Global Cascades," the difference between these regions being the absolute or relative (i.e., it scales with network size) minimum seed size, respectively, that is needed. A mechanism for the propagation of these TNCs is inherent in the theory. TNCs display the sigmoid shape typically observed and predicted in the theory of adoption of innovations and other growth theories.

In the Global Cascades region, the theory correctly predicts that a seed comprising a single node will launch a TNC, consistent with the infinite network theory except that TNCs are predicted to occur somewhat inside the No Global Cascades region, consistent with simulations. The theory correctly predicts that these processes are expanding.

In the No Global Cascades region, the same theory correctly predicts that TNCs require a relative minimum "critical mass" or "quorum" of seed size in order to emerge. TNCs in this region, if they occur, are correctly predicted to start out non-expanding unless the seed size is well above the transition value. We also showed by simulation that the minimum seed size in the No Global Cascades region displays a sharp phase transition and that the theory can predict with good accuracy the middle or top of the range of seed size over which this transition occurs. This allows us to predict with good accuracy a size of seed that is surely large enough to cause a TNC for a given network size, average nodal degree, and threshold. In addition, if the seed has an intermediate relative size such that TNCs emerge in simulations with probability less than one, a second critical mass is observed, comprising the nodes that need only one more flipped neighbor in order to flip. We showed how to calculate the minimum size of this second cohort and showed that tentative cascades survive or die depending on whether this minimum is exceeded or not.

We also showed that ever-increasing flipped cohort size is a sufficient condition for a cascade to emerge, since for intermediate values of $|S|$ a cascade can begin by flipping fewer nodes on each step and then reverse and become a TNC. The size criterion for the number one short can be seen as a tighter sufficient condition for the emergence of TNCs that are not expanding from the first step.

In order to obtain these results, we had to take account of the fact that the degree sequences of flipped and unflipped nodes are not binomial after the first step and that their respective average nodal degrees diverge from that of the original network. In addition, we had to calculate the number of incoming and outgoing edges explicitly after each step as functions of $K*$ and each node's degree.

Combining the present results with those in Whitney [16], we can identify three phase transitions in simulations of the threshold cascade problem. The first occurs as $z$ first exceeds unity. The second occurs a bit below the theoretical upper phase boundary identified by Watts, where TNCs occur because the process escapes the vulnerable clusters via the mechanism of cluster-hopping. The third occurs a bit above the theoretical upper phase boundary where processes need seed sizes larger than unity, sizes that scale with network size. These phase boundaries can be represented as defining regions in Figure 15. The theory presented here can distinguish two of these regions. The boundaries of Region II are not sharp and depend on the size of vulnerable clusters. No theory presently exists to define the required sizes or the exact extent of this region.

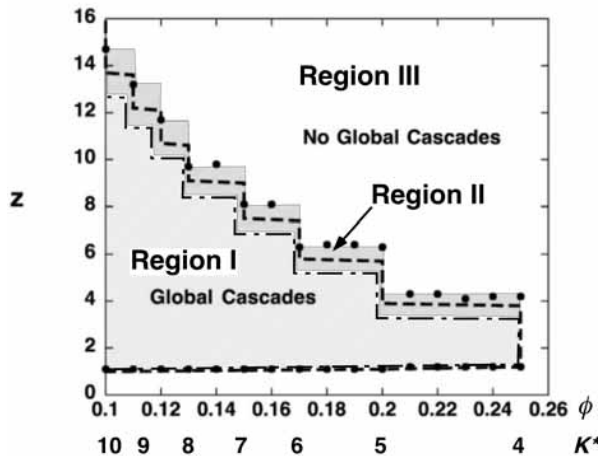

Figure 15. Phase Boundaries and Corresponding Regions. In Region I, TNCs comprise one huge vulnerable cluster or the process begins in a large vulnerable cluster and subsequently surrounds and flips stable nodes. These cascades are expanding. In Region II, cluster-hopping allows processes to become TNCs even though vulnerable nodes are few and vulnerable clusters are less than 1% of the network in size. These cascades are also expanding, but they are rare. In Regions I and II, a single node seed can start a TNC. In Region III, seeds must be of sufficient size, scaling with network size, in order for TNCs to occur. For seeds of transitional size, these cascades are non-expanding while for seeds larger than transitional size the processes are expanding. The theory in this paper combines Regions I and II and can distinguish this combined region from Region III.

The theory has several limitations. First, while it can calculate the degree sequence of newly flipped nodes, this information is not used completely. Instead, only the size and average nodal degree of the flipped set is used, and a binomial distribution with this average nodal degree is substituted for the real distribution.

Second, the theory does not recognize whether or not the network is connected. It thus fails to flip the whole network when the simulation easily does so. It also does not know of the existence of separate clusters of vulnerable nodes so it cannot reproduce cluster-hopping. Thus it requires slightly larger seeds in the upper boundary region of Figure 1 than are needed by the simulations. In retrospect, we can see that seed size $|S|=1$ never really occurs in the simulations in this region because, if the seed node links to a vulnerable cluster, the seed effectively enlarges to the size of that cluster (or clusters). Since the clusters comprise nodes with relatively small nodal degree, their ability to expand the seed is reduced by the ratio of the average nodal degree of vulnerable clusters (typically 80 to 90% of $K*$)[12] to the average nodal degree of the network at large. Nevertheless, the clusters multiply the effect of the seed. In fact, for every triplet of $z$, $K*$, and $S$ in the No Global Cascades region, the effective value of $|S|$ is greater than unity.

The approach taken here and in Whitney [16] relies on careful examination of the step-by-step evolution of individual simulations and presents a theory that captures that evolution probabilistically rather than seeking to predict on a static ensemble basis whether the network is likely to percolate, as past analyses have done.

The dynamic theory presented here should prove useful in understanding the behavior of highly-connected networks. It is more complex than the typical generating function approach and the calculations are tedious, but it makes more realistic assumptions about the nature of the network. Using it has enabled us to discover and study previously unrecognized behaviors, such as the "near death" and second critical mass phenomena.

---

[12] The average nodal degree of vulnerable clusters is not $z \approx 1$, as one might believe. $z \approx 1$ is the average nodal degree of the subgraph consisting of only the vulnerable nodes if all their links to nonvulnerable nodes are ignored. But their links to these other nodes are responsible for cluster-hopping and thus must be recognized and counted in order for this phenomenon to be understood.

## VI. Acknowledgements

The author thanks Duncan Watts, Peter Dodds, and Yaneer Bar'Yam for positive feedback and encouragement, and Professors Daniel Krob and Christophe Midler of l'Ecole Polytechnique de Paris for facilities, fruitful discussions, and financial support. The author especially thanks Raissa D'Souza for closely reading the manuscript and offering important suggestions and modifications.

## VII.    References

[1] T. Petermann and P. De Los Rios, Phys Rev E **69**, 066116 (2004).
[2] D. Centola, V. Eguiluz, and M. Macy, Physica A **374**, 449 (2007).
[3] M. E. J. Newman, Chapter 2 in S. Bornholdt and H. G. Schuster, eds., *Handbook of Graphs and Networks*, New York: Wiley, 2006
[4] D. S. Calloway, M. E. J. Newman, S.H. Strogatz, and D. J. Watts, Phys Rev Lett **85**, 5468 (2000).
[5] M. E. J. Newman, S. H. Strogatz, and D. J. Watts, *Phys. Rev. E* **64**, 026118 (2001)
[6] M. Molloy and B. Reed, Rand Struct Algorithms **6**, 161 (1995).
[7] D. J. Watts, "A Simple Model of Global Cascades on Random Networks," *PNAS* **99** 5766, 30 April 2002
[8] D. Lopez-Pintado, Games and Economic Behavior **62**, 573-590 March 2008,
[9] M. E. J. Newman, Phys. Rev. E **66**, 016128 (2002).
[10] E., J. Volz, Math. Biol. (2008) **56**:293–310
[11] D. Lopez-Pintado, Int J Game Th **34**, 371 (2006).
[12] T. Tlusty, and J.-P. Eckmann. J Phys A **42** 205004 (2009)
[13] M. O. Jackson, and L. Yariv, Economie Publique, **16** 1 3-16 (2005)
[14] Gleeson, J. P., and Cahalane, D. J., Phys. Rev. E. **75**, 056103 (2007)
[15] Gleeson, J. P., Phys. Rev. E. **77**, 046117 (2008)
[16] D. E. Whitney, "Exploring Watts' Cascade Boundary," International Conference on Complex Systems, Boston, November 2007. Accessible at http://www.interjournal.org/ manuscript number 2008
[17] F. Bass, (1969). "A new product growth model for consumer durables". *Management Science* **15** (5): p215–227
[18] Z. Griliches, (1957) "Hybrid Corn: An Exploration in the Economics of Technological Change," *Econometrica* vol. 25 (October).
[19] E. M. Rogers, *Diffusion of Innovations* (Free Press, New York, 2003)
[20] T. W. Valente, *Network Models of the Diffusion of Innovations*, Hampton Press, 1995
[21] R. Cohen, K. Erez, D. ben-Avraham, and S. Havlin, Phys Rev Lett **85** 21 4626-4628 (2000)

## VIII.    Appendix: Calculating $p_{nSF1}$

The network as a whole contains $n$ nodes connected with probability $p$. The network is originally comprised of two sets, the seed $S$ and the remaining nodes $n-S$. The seed nodes are flipped to start the process, and the remaining nodes are initially unflipped. The seed nodes flip a new set called $F1$, leaving an unflipped set comprising $n-S-F1$ nodes. $p_{nSF1}$ is the probability that two nodes among this unflipped set are linked. We anticipate that $p_{nSF1} > p$ because this unflipped set is biased toward having more edges than the original network because it is depleted by $F1$, a set of nodes that has fewer edges than average. This trend continues for subsequent steps.

To start the calculation, we note that the nodes in $n-S$ have $z(n-S)$ edges emanating from them[13], including edges shared with $S$. Since edges are connected at random, the share of these edges that connect only the nodes in $n-S$ to each other comprise a fraction $(n-S)/n$ of these edges. We define $z_{nS}$ to be the average nodal degree of nodes in $n-S$ considering only the edges that connect those nodes to each other (called internal edges). Internal edges are the correct ones to count because Equation 3 deals only with unflipped nodes and treats their links to each other as distinct from their links to $S$. Then we can calculate

**Equation 19**

$$internal\ edges = z(n-S)\frac{n-S}{n}$$

$$internal\ nodes = n-S$$

$$z_{nS} = \frac{internal\ edges}{internal\ nodes} = \frac{z(n-S)^2/n}{n-S} = \frac{z(n-S)}{n}$$

Similarly, once $F1$ nodes have flipped, there are $n-S-F1$ unflipped nodes and from these nodes emanate $z(n-S-\alpha_{F1}F1)$ edges, where $0 < \alpha_{F1} < 1$ is a fraction conveying the fact that $F1$ has fewer edges than average. $\alpha_{F1}$ is the ratio $zF1_0/z$, where $zF1_0$ is the average nodal degree of $F1$. Numerically, $\alpha_{F1} \approx 0.74$ for the case where $n = 4500, z = 11.5, K^* = 4$ for the first step, and has slightly larger values for the next several steps. Then we can calculate

---

[13] This formulation counts each edge twice.

**Equation 20**

$$\text{internal edges} = z(n - S - \alpha_{F1}F1)\frac{n - S - F1}{n}$$

$$\text{internal nodes} = n - S - F1$$

$$z_{nSF1} = \frac{\text{internal edges}}{\text{internal nodes}} = \frac{z(n - S - \alpha_{F1}F1)(n - S - F1)/n}{(n - S - F1)} = \frac{z(n - S - \alpha_{F1}F1)}{n}$$

And, using similar logic, we have

**Equation 21**

$$z_{nSF1F2} = \frac{z(n - S - (\alpha_{F1}F1 + \alpha_{F2}F2))}{n}$$

Subsequent steps follow the same pattern.

To calculate $p_{nSF1}$ directly, we use

# Equation 22

$$p_{nSF1} = \frac{2 * \text{internal edges}}{\text{internal nodes}^2} = \frac{z(n - S - \alpha_{F1}F1)(n - S - F1)/n}{(n - S - F1)^2}$$

$$= \frac{z(n - S - \alpha_{F1}F1)}{n(n - S - F1)} = p\frac{(n - S - \alpha_{F1}F1)}{(n - S - F1)}$$